\documentclass[11pt,english,notitlepage, onecolumn]{revtex4-1}
\usepackage[T1]{fontenc}
\usepackage[scaled]{helvet}
\usepackage{color}
\usepackage{amsmath}
\usepackage{amssymb}
\usepackage{graphicx}

\usepackage{babel}
\begin{document}
	
	\title{Broad frequency shift of parametric processes in Epsilon-Near-Zero time-varying media}
	
\author{Vincenzo Bruno $^{1}$, Stefano Vezzoli $^{2}$, Clayton DeVault$^{3,4}$, Enrico Carnemolla$^{5}$, Marcello Ferrera$^{5}$, Alexandra Boltasseva$^{3,4}$, Vladimir M. Shalaev$^{3,4}$, Daniele Faccio $^{1, \dagger}$ and Matteo Clerici$^{6, \dagger}$}
	
	\affiliation{$^1$ School of Physics and Astronomy, University of Glasgow,  G12 8QQ Glasgow, United Kingdom}
	\affiliation{$^2$ The Blackett Laboratory, Department of Physics, Imperial College London, London SW7 2BW, United Kingdom}
	\affiliation{$^3$ Purdue Quantum Science and Engineering Institute, Purdue University 1205 West State Street, West Lafayette, Indiana 47907, USA}
	\affiliation{$^4$ School of Electrical and Computer Engineering and Birck Nanotechnology Center, Purdue University, 1205 West State Street, West Lafayette, Indiana 47907, USA}
	\affiliation{$^5$ Institute of Photonics and Quantum Sciences, Heriot-Watt University,  EH14 4AS Edinburgh, United Kingdom}
	\affiliation{$^6$ School of Engineering, University of Glasgow, G12 8LT Glasgow, United Kingdom}
	\email{ $^{\dagger}$ daniele.faccio@glasgow.ac.uk, $^{\dagger}$matteo.clerici@glasgow.ac.uk.}
	
	\begin{abstract}
		$\mathbf{Abstract:}$~The ultrafast changes of material properties induced by short laser pulses can lead to frequency shift of reflected and transmitted radiation. Recent reports highlight how such a frequency shift is enhanced in the spectral regions where the material features a near-zero real part of the permittivity. Here we investigate the frequency shift for fields generated by four-wave mixing with a nonlinear polarisation oscillating at twice the pump frequency. In our experiment we observe a frequency shift of more than $60$~nm (compared to the pulse width of $\sim$40 nm) for the phase conjugated radiation generated by a $500$~nm Aluminium-doped Zinc Oxide (AZO) film pumped close to the epsilon-near-zero wavelength. Our results indicate applications of time-varying media for nonlinear optics and frequency conversion.
	\end{abstract}
	
	\maketitle

$\mathbf{Keyword}$~Epsilon-near-zero media; four-wave mixing; Frequency shift; Transparent conductive oxides; Optical Kerr nonlinearities; time-varying media.








\setcounter{section}{0} 
\section{Introduction}
Controlling the phase of an optical signal with a temporal resolution below one picosecond could prove essential for ultrafast signal processing. Such short time-scales can be achieved exploiting nonlinear light-matter interaction in second order nonlinear crystals (electro-optical effect) or Kerr effects in centrosymmetric media. In these processes, the instantaneous frequency is temporally modulated, and if the temporal phase change is uniform over the pulse duration, the optical spectrum can be rigidly shifted. This effect can be interpreted as time-refraction \cite{mendoncca2000theory,auyeung1983PCtemporal}, a phenomenon that attracted the attention of the research community owing to its link with the dynamical Casimir effect and Hawking radiation, and its implication in the formation of temporal band-gap structures and non reciprocal devices~\cite{mendonca2008vacuum, Faccio2014quantumcosmology, faccio2011dynamical, shaltout2015time, martinez2016temporal, PRLprain}. 
Strong enhancement of light-matter interaction has been observed in the spectral region where the real part of the dielectric permittivity ($\Re[\varepsilon_r]$) of a medium approaches zero (ENZ)~\cite{engheta2013pursuingENZ, luk2015ENZmodes, liberal2017ENZreview}. In the case of sub-wavelength thin films of transparent conductive oxide (TCOs), such enhancement combined with an ultrafast response of the medium~\cite{capretti2015TGHITOENZ,kinsey2015ENZlinear,alam2016ENZnonlinear, PRLCaspani,clericiNC,yang2017femtosecond,carnemolla2018degenerate,niu2018,wood2018,boyd2019NatRev,amr2019} results in the generation of phase conjugation (${\text{PC}}$) and negative refraction (${\text{NR}}$) with near unity conversion efficiency~\cite{VezzoliTime}. One of the intriguing effects associated with the enhanced ultrafast material response is the observation of spectral shifts of the order of the pulse bandwidth ($10-15$~nm) for the transmitted and the reflected radiation in pump-and-probe experiments~\cite{PRLCaspani, clericiNC}. This phenomenon has been interpreted as originating from the adiabatic shift of the material refractive index enhanced, at the ENZ wavelength, by the associated slow-light condition~\cite{khurgin2019}. 

In this work, we investigate the spectral shift of PC and NR in a time-varying TCO film (Aluminium-Doped Zinc-Oxide (AZO)~\cite{kinsey2015ENZlinear}) at the ENZ wavelength. We report the efficient generation of a $>60$~nm wavelength red-shifted PC and of a both red and blue-shifted NR (covering nearly the same bandwidth, $\simeq60$~nm) from a $500$~nm thick AZO film pumped at the ENZ wavelength. 
 
 \begin{figure}[h!]
 	\centering
 	\fbox{\includegraphics[width=\linewidth]{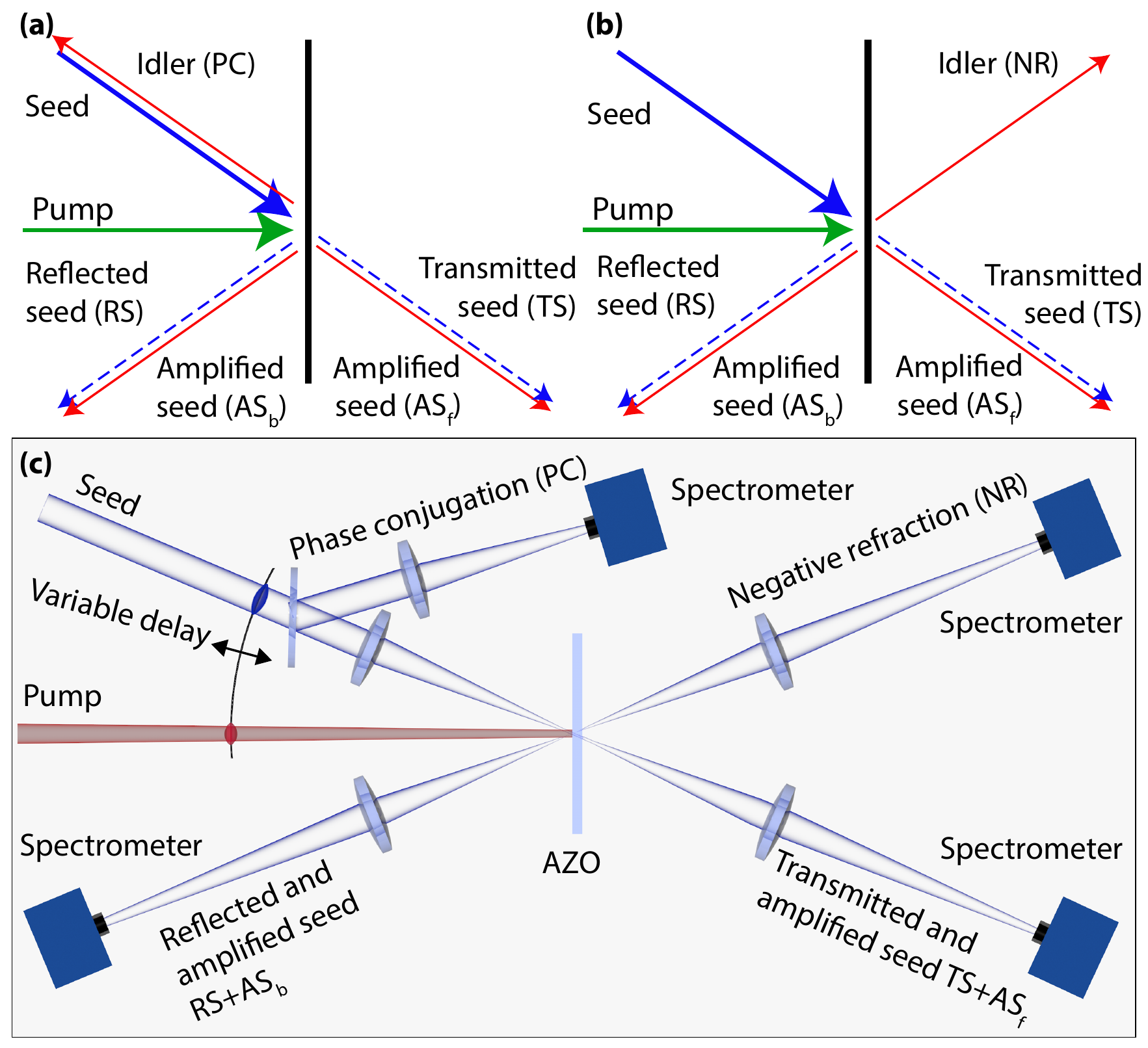}}
 	\caption{ {\bf (a)}  and {\bf (b)} describe the two four-wave mixing processes giving birth to the PC an NR signals, respectively. The pump beam is illustrated with a green arrow while the input seed is shown with a blu arrow. Red arrows show the result of the parametric amplification process while blue dashed arrows show the linear reflection and transmission. For each of these two four-wave mixing processes, the seed field is amplified and emerges along both the transmitted and reflected directions. {\bf (c)} Sketch of the experimental setup. Pump and seed fields have the same wavelength ($1400$~nm) and vertical polarization. The seed is overlapped to the pump with an angle of ${\simeq6}$~deg.  The delay between the pump and seed pulses is controlled by a delay line (not shown). A spectrometer is used to collect the radiation emerging along the direction of the seed reflection and transmission ($\text{RS}+\text{AS}_{\text{b}}$ and $\text{TS}+\text{AS}_{\text{f}}$), as well as to characterize the PC and NR.}
 	\label{fig:scheme}
 \end{figure}

\section{Results}
Phase conjugation and negative refraction result from a four-wave mixing interaction mediated by a Kerr-type nonlinearity and satisfy a parametric amplification energy matching condition: $\omega_{\text{PC, NR}}=2\omega_{\text{P}}-\omega_{\text{S}}$, where $\omega_{\text{p}}$ and $\omega_{\text{s}}$ are the pump ($\text{P}$) and seed ($\text{S}$) frequencies, respectively~\cite{pendry2008time, PhysRevAPendry, palomba2012optical, rao2015geometries}. The PC and NR are the processes satisfying the additional condition $\omega_{\text{S}}=\omega_{\text{P}}=\omega_0$, leading to $\omega_{\text{PC, NR}}=\omega_0$. We remind that for each photon of NR and PC generated as \emph{idler} of the parametric amplification process, a \emph{seed} photon is also amplified (amplified seed - $\text{AS}$). In the case of a deeply sub-wavelength nonlinear medium, the component of momentum in the direction of the film thickness ($k_z$) does not need to be conserved, and the only relevant phase matching condition involves the in-plane wavevector components $\vec{k}_r$, according to:
\begin{equation}
\vec{k}_{r}^{(\text{NR,PC})}+\vec{k}_{r}^{(\text{AS})}=2\,\vec{k}_{r}^{(\text{P})}.
\end{equation}
Since the pump beam is orthogonal to the nonlinear film, it has a null in-plane wavevector component ($k_r^{(\text{P})}=0$), and the above constraint simplifies to $\vec{k}_{r}^{(\text{NR,PC})}=-\vec{k}_{r}^{(\text{AS})}$.
The amplification process is coherent and the amplified seed retains the phase of the input seed. Due to the sub-wavelength nature of the nonlinear medium, the longitudinal component of the amplified seed wavevector is undetermined, and the only spatial requirement is that $\vec{k}_{r}^{(\text{AS})}=\vec{k}_{r}^{(\text{S})}$, which has a forward and a backward solution, as shown in Figs.~\ref{fig:scheme}~(a) and (b). In summary, for a seed photon injected into the nonlinear medium,  one photon will emerge either as the phase conjugation (${\text{PC}}$) or negative refraction  (${\text{NR}}$) of the input seed. At the same time another photon (amplified seed) will be generated, either in the backward ($\text{AS}_{\text{b}}$) or in the forward ($\text{AS}_{\text{f}}$) direction. These last two signals are overlapped to the reflected ($\text{RS}$) and the transmitted ($\text{TS}$) seed, respectively. We assume that pump and seed are co-polarized. In the ideal case described here, the amplified seed can emerge either in the backward or in the forward direction, irrespective from the generation of negative refraction or phase conjugation. We shall see below how the frequency shift of the $\text{PC}$ and $\text{NR}$ are linked to those of the $\text{RS}$ and $\text{TS}$ fields.

In previous experiments we showed that PC and NR generated by a sub-wavelength AZO film can achieve larger than unitary internal efficiency, meaning that the PC and NR have, inside the sample, an amplitude larger than the seed~\cite{VezzoliTime}. 
\begin{figure}[h]
	\centering
	\fbox{\includegraphics[width=\linewidth]{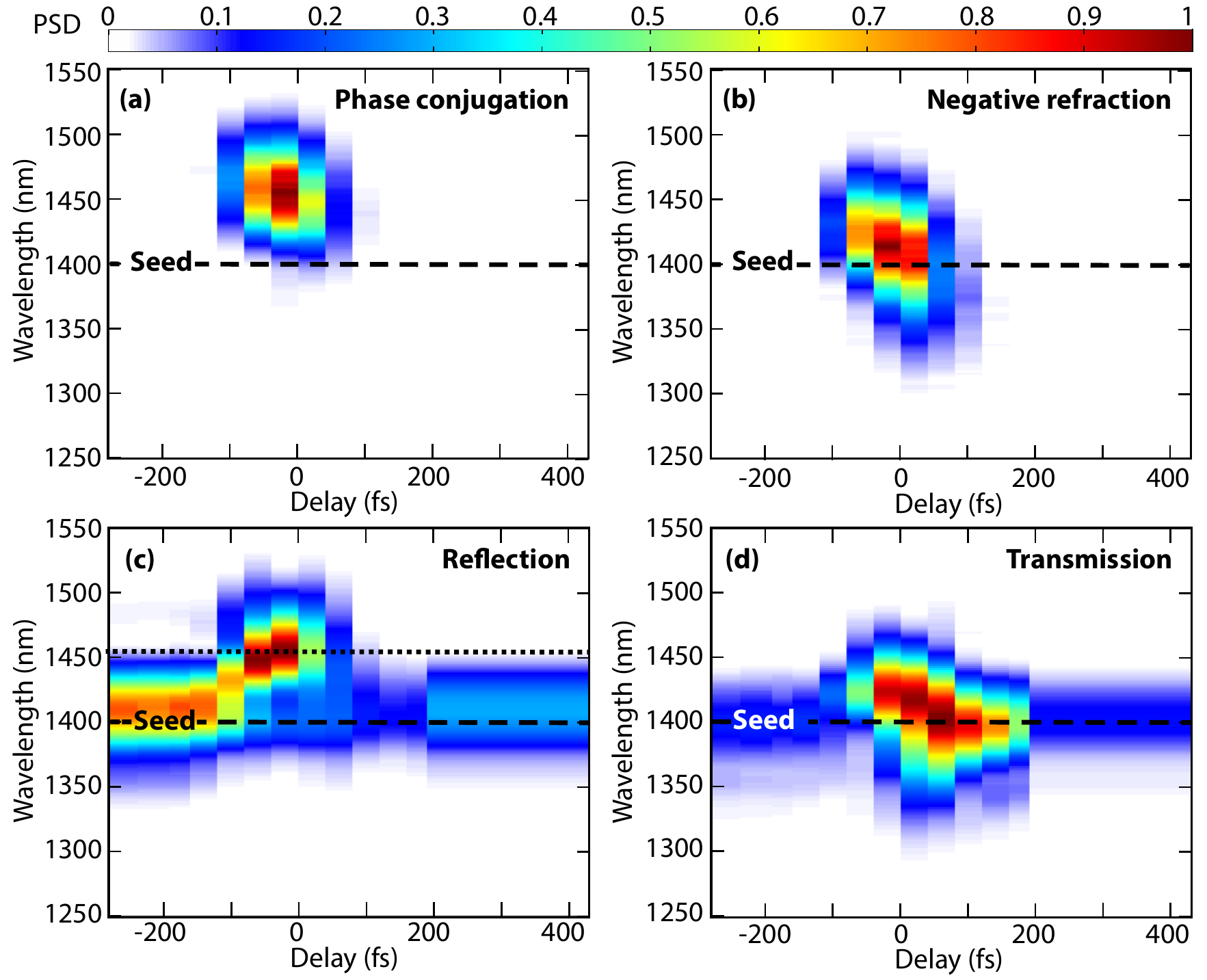}}
	\caption{Measured spectrograms (normalized Power Spectral Density - PSD) for {\bf (a)} the PC,  {\bf (b)} the NR,  {\bf (c)} the reflection, and  {\bf (d)} the transmission of the input seed field. The horizontal axis in each panel denotes the delay between the pump and the seed pulses. The vertical axis indicates the wavelength. The dashed line is at the seed carrier wavelength ($1400$~nm), while the dotted line refers to the amplified seed measurement (see text for details).}
	\label{fig:fig2}
\end{figure}
Here we investigate the spectral dynamics of PC, NR, and of the amplified seed, in similar experimental conditions. We performed the degenerate FWM experiment using a pump and probe set-up schematically shown in Fig.~\ref{fig:scheme}~(c). The output of a mode-locked Ti:sapphire laser with a pulse duration of $\simeq 105$~fs pumped an optical parametric amplifier (OPA), which generated a pulse train with wavelength tuneable in the near-IR spectral region. The OPA output was split in two arms. The pump, with high intensity, was injected at normal incidence onto the ENZ film. The seed, with a lower intensity and smaller beam spot size, was overlapped with the pump coming at a small ($\simeq 6^{\circ}$) angle to the sample normal. The $500$~nm AZO film was deposited on a glass substrate by laser pulse deposition and exhibits a real part of the relative dielectric permittivity $\Re[\varepsilon_{\text{r}}]$ crossing zero at $\lambda_{\text{ENZ}} \simeq 1350$~nm. We choose to excite the AZO film above the crossing point ($\lambda_{\text{p}}=\lambda_{\text{s}}=1400$~nm). This choice was motivated by recent findings showing that the frequency degenerate FWM is more efficient at wavelengths longer than the $\lambda_{ENZ}$~\cite{VezzoliTime,carnemolla2018degenerate}. For the wavelength of choice, the real and imaginary part of the refractive index of the AZO sample measured by ellipsometry were $n_{\text{r}}\simeq 0.3$ and $n_\text{i}\simeq\, 0.89$, respectively. The pump intensity was set to $769$~GW/cm$^{2}$. 
We recorded the power spectra of PC (Fig.~\ref{fig:fig2}~(a)), NR (b), reflection (c), and transmission (d) as a function of the delay between the pump and the seed ($\tau$, where $\tau<0$ for the seed impinging on the sample before the pump pulse). The spectra of the nonlinear signals were collected using the same spectrometer. The normalized spectrograms show that at the delay corresponding to maximum generation efficiency, both the PC and NR are red-shifted with respect to the input seed, yet with different spectral shifts. The transmission and reflection spectrograms have a more complex structure that we interpret as the overlap of the signal reflection and transmission with the amplified seed field ($\text{AS}_{\text{b}}$ and $\text{AS}_{\text{f}}$ respectively).

\begin{figure}[h]
	\centering
	\fbox{\includegraphics[width=\linewidth]{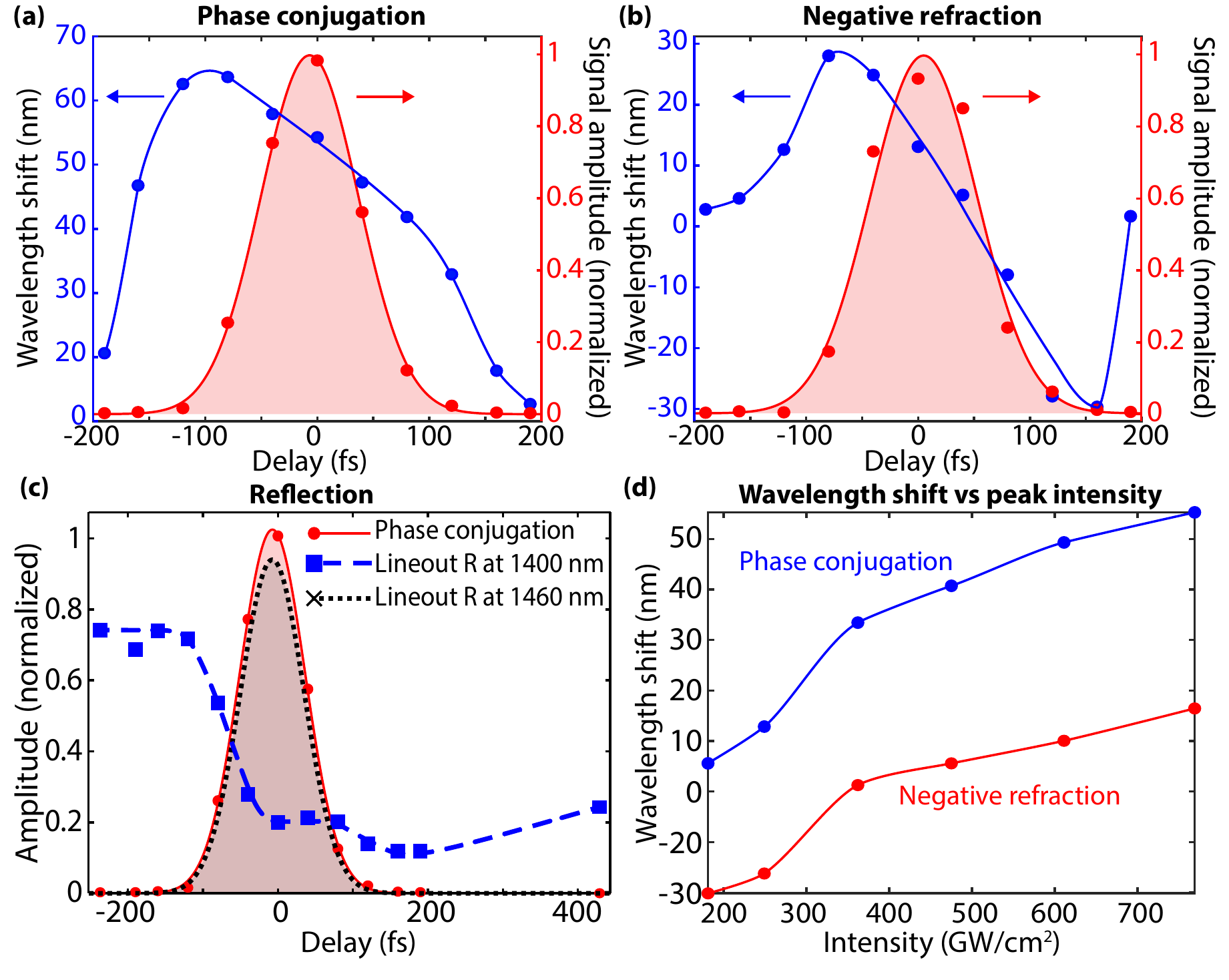}}
	\caption{(a) Carrier wavelength shift of the PC signal as a function of the pump-probe delay $\Delta\tau$ (blue-curve) overlapped with the PC signal amplitude (red curve). (b) Carrier wavelength shift of the NR signal as a function of the pump-probe delay $\Delta\tau$ (blue curve) overlapped with the PC signal amplitude (red curve). (c) Reflection spectrogram as in Fig.~\ref{fig:fig2}(c), with a dashed blue line at the input seed carrier wavelength and a black dotted line at the carrier wavelength of the PC peak. (d) Measured wavelength shifts for the PC (blue) and the NR (red) nonlinear products, as a function of the pump intensity.}
	\label{fig:fig3}
\end{figure}
To better understand the spectral evolution, we plot the carrier wavelength shift (blue line) as a function of the pump-probe delay for both the PC and NR signals in Figs.~\ref{fig:fig3}~(a) and (b), respectively. We overlap the wavelength shift with the normalized energy (red) of the fields. One can see that the PC is shifted by more than $60$~nm with respect to the input seed  wavelength, with a maximum shift occurring for the seed overlapped with the leading edge of the pump pulse. Conversely, the NR wavelength shift is larger when the seed is overlapped with either the leading or the trailing edge of the pump pulse, yet with the opposite sign. The reflected seed is expected to drop and slowly regain the original unperturbed wavelength as a function of the pump-probe delay $\Delta\tau$, as shown for an intraband excitation in Ref.~\cite{PRLCaspani}. This explains, for instance, why the reflection at $-200$~fs is stronger than that at $400$~fs. Unlike Ref.~\cite{PRLCaspani}, after an initial drop the reflected signal suddenly increases at a wavelength $\simeq 60$~nm red-shifted with respect to the input seed, as shown in Fig.~\ref{fig:fig3}(c). We interpret this as the combination of the reflection dynamics already reported in~\cite{PRLCaspani} with the signal generated by the parametric process in conjunction to PC and NR. We note that this shift is larger than the probe spectral bandwidth of the input seed ($\simeq40$~nm). A similar situation can be qualitatively observed along the transmission direction. However, since the frequency shift for NR switches from positive to negative it is impossible to completely separate the two contributions in Fig.~\ref{fig:fig2}(d).

\section{Discussion}
The broad frequency shift observed for four-wave mixing around the epsilon-near-zero region only partially arises from the same physical mechanism responsible for the frequency shift of reflected and transmitted beams from a time-varying surface~\cite{khurgin2019}.
The fields generated by the nonlinear polarization experience an adiabatic shift of the refractive index. However a simple application of that model to our experiment would predict shifts comparable to those reported in~\cite{khurgin2019}, which are nearly three times smaller than what is reported here. 
Another contribution to the frequency shift of all beams originates from the time-varying Fresnel coefficients at the two interfaces air/AZO and AZO/glass. A complex refractive index that varies in time at a faster-than-picosecond time scale, as in our case, lead to significant changes in the wavelength of the impinging fields. We have estimated the magnitude of the effect that such a change would provide assuming a large change of refractive index ($\Delta n=2$), delivered in a short, $50$~fs time period. Furthermore, even for this extreme case the time dependence of the complex Fresnel coefficients would only account for $12$~nm of shift in NR and $35$~nm in PC. Furthermore, even combining the two effects we could only account for about 35~$\%$ of the observed shift for the forward PC. We therefore speculate that the experimental results indicate possibly non-trivial underlying physical effects that might require a full microscopic model of this experimental study.

Finally, we note that the simple model of an infinitely thin time-varying medium presented in Fig.~\ref{fig:scheme}(a) and (b) would imply a symmetric effect in the forward and backward directions. In contrast, we observe that PC and NR have different spectral dynamics, and each of them matches that of the amplified seed along the reflection (PC) and transmission (NR). Overall, the forward signals generated by the four-wave mixing process show a spectral blue shift with respect to those generated in the backward direction, as seen in Fig.~\ref{fig:fig3}(d).  This figure shows the dependence of the wavelength shift for PC (blue) and NR (red) for increasing pump intensities. The two curves follow a qualitatively similar trend, yet with a rigid blue shift of the NR that is $30$~nm larger. The asymmetry between forward and backward directions may be assigned to the difference between the Fresnel coefficients experienced by the forward and backward propagating beams. Absorption, which induces an exponential decay of all beams (the attenuation coefficient is $\alpha=4\pi n_{i}/\lambda=0.0079~\text{nm}^{-1}$), may also contribute to the observed asymmetries, as well as the cross phase modulation between the pump and the generated radiation, which mainly affects the forward propagating beam. Once again, the simple models outlined above can explain some of the difference between backward and forward, but intriguingly cannot fully account for the observed frequency shift.   

\section{Conclusions}

In conclusion, we have experimentally investigated the frequency shift for PC, NR, and for the amplified signals generated by a four-wave mixing process in a thin-film of AZO in the ENZ wavelength region. We observed extremely large wavelength shifts, exceeding $60$~nm in the backward propagating radiation. We also recorded a different wavelength shift for the forward and backward scattered waves. Attempts to explain the observations with recently proposed models are not fully satisfactory, implying that the dynamics of nonlinear time-varying media still requires further theoretical investigation. Our results raise new questions about the fundamental physics of time-varying media in the ENZ regime and pave the way to a new range of applications in integrated nonlinear optics, e.g. for frequency conversion and optical switch. All the data supporting this manuscript are available at the DOI: xx/xxx/xx.

\vspace{6pt}

$\mathbf{Authorcontributions}$~V.B., S.V., E.C. acquired the data, C.D. fabricated the sample. All authors contributed to the conceptualization of the work; V.B., S.V., D.F., and M.C. contributed to the formal analysis; M.F., A.B., V. S., D.F. and M.C. provided the required resources; V.B., S.V., D.F. and M.C. contributed to the original draft preparation; all the authors contributed to writing, review and editing of the manuscript.

$\mathbf{Funding}$~M.C. acknowledges the funding from UK Research and Innovation, \emph{In-Tempo}, Innovation Fellowship EP/S001573/1. DF acknowledges financial support from EPSRC (UK Grants EP/M009122/1 and EP/P006078/2). Purdue	team acknowledges support by the U.S. Department of Energy, Office of Basic Energy Sciences, Division of Materials Sciences and Engineering under Award DE-SC0017717 (sample preparation), Air Force Office of Scientic Research (AFOSR) award 	FA9550-18-1-0002 and Office of Naval Research (optical characterization).


\begin{thebibliography}{28}%
	\makeatletter
	\providecommand \@ifxundefined [1]{%
		\@ifx{#1\undefined}
	}%
	\providecommand \@ifnum [1]{%
		\ifnum #1\expandafter \@firstoftwo
		\else \expandafter \@secondoftwo
		\fi
	}%
	\providecommand \@ifx [1]{%
		\ifx #1\expandafter \@firstoftwo
		\else \expandafter \@secondoftwo
		\fi
	}%
	\providecommand \natexlab [1]{#1}%
	\providecommand \enquote  [1]{``#1''}%
	\providecommand \bibnamefont  [1]{#1}%
	\providecommand \bibfnamefont [1]{#1}%
	\providecommand \citenamefont [1]{#1}%
	\providecommand \href@noop [0]{\@secondoftwo}%
	\providecommand \href [0]{\begingroup \@sanitize@url \@href}%
	\providecommand \@href[1]{\@@startlink{#1}\@@href}%
	\providecommand \@@href[1]{\endgroup#1\@@endlink}%
	\providecommand \@sanitize@url [0]{\catcode `\\12\catcode `\$12\catcode
		`\&12\catcode `\#12\catcode `\^12\catcode `\_12\catcode `\%12\relax}%
	\providecommand \@@startlink[1]{}%
	\providecommand \@@endlink[0]{}%
	\providecommand \url  [0]{\begingroup\@sanitize@url \@url }%
	\providecommand \@url [1]{\endgroup\@href {#1}{\urlprefix }}%
	\providecommand \urlprefix  [0]{URL }%
	\providecommand \Eprint [0]{\href }%
	\providecommand \doibase [0]{http://dx.doi.org/}%
	\providecommand \selectlanguage [0]{\@gobble}%
	\providecommand \bibinfo  [0]{\@secondoftwo}%
	\providecommand \bibfield  [0]{\@secondoftwo}%
	\providecommand \translation [1]{[#1]}%
	\providecommand \BibitemOpen [0]{}%
	\providecommand \bibitemStop [0]{}%
	\providecommand \bibitemNoStop [0]{.\EOS\space}%
	\providecommand \EOS [0]{\spacefactor3000\relax}%
	\providecommand \BibitemShut  [1]{\csname bibitem#1\endcsname}%
	\let\auto@bib@innerbib\@empty
	\bibitem [{\citenamefont {Mendon{\c{c}}a}(2000)}]{mendoncca2000theory}%
	\BibitemOpen
	\bibfield  {author} {\bibinfo {author} {\bibfnamefont {J.~T.}\ \bibnamefont
			{Mendon{\c{c}}a}},\ }\href@noop {} {\emph {\bibinfo {title} {Theory of photon
				acceleration}}}\ (\bibinfo  {publisher} {CRC Press},\ \bibinfo {year}
	{2000})\BibitemShut {NoStop}%
	\bibitem [{\citenamefont {AuYeung}(1983)}]{auyeung1983PCtemporal}%
	\BibitemOpen
	\bibfield  {author} {\bibinfo {author} {\bibfnamefont {J.~C.}\ \bibnamefont
			{AuYeung}},\ }\href@noop {} {\bibfield  {journal} {\bibinfo  {journal}
			{Optics letters}\ }\textbf {\bibinfo {volume} {8}},\ \bibinfo {pages} {148}
		(\bibinfo {year} {1983})}\BibitemShut {NoStop}%
	\bibitem [{\citenamefont {Mendonca}\ \emph {et~al.}(2008)\citenamefont
		{Mendonca}, \citenamefont {Brodin},\ and\ \citenamefont
		{Marklund}}]{mendonca2008vacuum}%
	\BibitemOpen
	\bibfield  {author} {\bibinfo {author} {\bibfnamefont {J.~T.}\ \bibnamefont
			{Mendonca}}, \bibinfo {author} {\bibfnamefont {G.}~\bibnamefont {Brodin}}, \
		and\ \bibinfo {author} {\bibfnamefont {M.}~\bibnamefont {Marklund}},\
	}\href@noop {} {\bibfield  {journal} {\bibinfo  {journal} {Physics Letters
				A}\ }\textbf {\bibinfo {volume} {372}},\ \bibinfo {pages} {5621} (\bibinfo
		{year} {2008})}\BibitemShut {NoStop}%
	\bibitem [{\citenamefont {Westerberg}\ \emph {et~al.}(2014)\citenamefont
		{Westerberg}, \citenamefont {Cacciatori}, \citenamefont {Belgiorno},
		\citenamefont {Dalla~Piazza},\ and\ \citenamefont
		{Faccio}}]{Faccio2014quantumcosmology}%
	\BibitemOpen
	\bibfield  {author} {\bibinfo {author} {\bibfnamefont {N.}~\bibnamefont
			{Westerberg}}, \bibinfo {author} {\bibfnamefont {S.}~\bibnamefont
			{Cacciatori}}, \bibinfo {author} {\bibfnamefont {F.}~\bibnamefont
			{Belgiorno}}, \bibinfo {author} {\bibfnamefont {F.}~\bibnamefont
			{Dalla~Piazza}}, \ and\ \bibinfo {author} {\bibfnamefont {D.}~\bibnamefont
			{Faccio}},\ }\href@noop {} {\bibfield  {journal} {\bibinfo  {journal} {New
				Journal of Physics}\ }\textbf {\bibinfo {volume} {16}},\ \bibinfo {pages}
		{075003} (\bibinfo {year} {2014})}\BibitemShut {NoStop}%
	\bibitem [{\citenamefont {Faccio}\ and\ \citenamefont
		{Carusotto}(2011)}]{faccio2011dynamical}%
	\BibitemOpen
	\bibfield  {author} {\bibinfo {author} {\bibfnamefont {D.}~\bibnamefont
			{Faccio}}\ and\ \bibinfo {author} {\bibfnamefont {I.}~\bibnamefont
			{Carusotto}},\ }\href@noop {} {\bibfield  {journal} {\bibinfo  {journal} {EPL
				(Europhysics Letters)}\ }\textbf {\bibinfo {volume} {96}},\ \bibinfo {pages}
		{24006} (\bibinfo {year} {2011})}\BibitemShut {NoStop}%
	\bibitem [{\citenamefont {Shaltout}\ \emph {et~al.}(2015)\citenamefont
		{Shaltout}, \citenamefont {Kildishev},\ and\ \citenamefont
		{Shalaev}}]{shaltout2015time}%
	\BibitemOpen
	\bibfield  {author} {\bibinfo {author} {\bibfnamefont {A.}~\bibnamefont
			{Shaltout}}, \bibinfo {author} {\bibfnamefont {A.}~\bibnamefont {Kildishev}},
		\ and\ \bibinfo {author} {\bibfnamefont {V.}~\bibnamefont {Shalaev}},\
	}\href@noop {} {\bibfield  {journal} {\bibinfo  {journal} {Optical Materials
				Express}\ }\textbf {\bibinfo {volume} {5}},\ \bibinfo {pages} {2459}
		(\bibinfo {year} {2015})}\BibitemShut {NoStop}%
	\bibitem [{\citenamefont {Mart{\'\i}nez-Romero}\ \emph
		{et~al.}(2016)\citenamefont {Mart{\'\i}nez-Romero}, \citenamefont
		{Becerra-Fuentes},\ and\ \citenamefont {Halevi}}]{martinez2016temporal}%
	\BibitemOpen
	\bibfield  {author} {\bibinfo {author} {\bibfnamefont {J.~S.}\ \bibnamefont
			{Mart{\'\i}nez-Romero}}, \bibinfo {author} {\bibfnamefont {O.}~\bibnamefont
			{Becerra-Fuentes}}, \ and\ \bibinfo {author} {\bibfnamefont {P.}~\bibnamefont
			{Halevi}},\ }\href@noop {} {\bibfield  {journal} {\bibinfo  {journal}
			{Physical Review A}\ }\textbf {\bibinfo {volume} {93}},\ \bibinfo {pages}
		{063813} (\bibinfo {year} {2016})}\BibitemShut {NoStop}%
	\bibitem [{\citenamefont {Prain}\ \emph {et~al.}(2017)\citenamefont {Prain},
		\citenamefont {Vezzoli}, \citenamefont {Westerberg}, \citenamefont {Roger},\
		and\ \citenamefont {Faccio}}]{PRLprain}%
	\BibitemOpen
	\bibfield  {author} {\bibinfo {author} {\bibfnamefont {A.}~\bibnamefont
			{Prain}}, \bibinfo {author} {\bibfnamefont {S.}~\bibnamefont {Vezzoli}},
		\bibinfo {author} {\bibfnamefont {N.}~\bibnamefont {Westerberg}}, \bibinfo
		{author} {\bibfnamefont {T.}~\bibnamefont {Roger}}, \ and\ \bibinfo {author}
		{\bibfnamefont {D.}~\bibnamefont {Faccio}},\ }\href {\doibase
		10.1103/PhysRevLett.118.133904} {\bibfield  {journal} {\bibinfo  {journal}
			{Phys. Rev. Lett.}\ }\textbf {\bibinfo {volume} {118}},\ \bibinfo {pages}
		{133904} (\bibinfo {year} {2017})}\BibitemShut {NoStop}%
	\bibitem [{\citenamefont {Engheta}(2013)}]{engheta2013pursuingENZ}%
	\BibitemOpen
	\bibfield  {author} {\bibinfo {author} {\bibfnamefont {N.}~\bibnamefont
			{Engheta}},\ }\href@noop {} {\bibfield  {journal} {\bibinfo  {journal}
			{Science}\ }\textbf {\bibinfo {volume} {340}},\ \bibinfo {pages} {286}
		(\bibinfo {year} {2013})}\BibitemShut {NoStop}%
	\bibitem [{\citenamefont {Luk}\ \emph {et~al.}(2015)\citenamefont {Luk},
		\citenamefont {De~Ceglia}, \citenamefont {Liu}, \citenamefont {Keeler},
		\citenamefont {Prasankumar}, \citenamefont {Vincenti}, \citenamefont
		{Scalora}, \citenamefont {Sinclair},\ and\ \citenamefont
		{Campione}}]{luk2015ENZmodes}%
	\BibitemOpen
	\bibfield  {author} {\bibinfo {author} {\bibfnamefont {T.~S.}\ \bibnamefont
			{Luk}}, \bibinfo {author} {\bibfnamefont {D.}~\bibnamefont {De~Ceglia}},
		\bibinfo {author} {\bibfnamefont {S.}~\bibnamefont {Liu}}, \bibinfo {author}
		{\bibfnamefont {G.~A.}\ \bibnamefont {Keeler}}, \bibinfo {author}
		{\bibfnamefont {R.~P.}\ \bibnamefont {Prasankumar}}, \bibinfo {author}
		{\bibfnamefont {M.~A.}\ \bibnamefont {Vincenti}}, \bibinfo {author}
		{\bibfnamefont {M.}~\bibnamefont {Scalora}}, \bibinfo {author} {\bibfnamefont
			{M.~B.}\ \bibnamefont {Sinclair}}, \ and\ \bibinfo {author} {\bibfnamefont
			{S.}~\bibnamefont {Campione}},\ }\href@noop {} {\bibfield  {journal}
		{\bibinfo  {journal} {Applied Physics Letters}\ }\textbf {\bibinfo {volume}
			{106}},\ \bibinfo {pages} {151103} (\bibinfo {year} {2015})}\BibitemShut
	{NoStop}%
	\bibitem [{\citenamefont {Liberal}\ and\ \citenamefont
		{Engheta}(2017)}]{liberal2017ENZreview}%
	\BibitemOpen
	\bibfield  {author} {\bibinfo {author} {\bibfnamefont {I.}~\bibnamefont
			{Liberal}}\ and\ \bibinfo {author} {\bibfnamefont {N.}~\bibnamefont
			{Engheta}},\ }\href@noop {} {\bibfield  {journal} {\bibinfo  {journal}
			{Nature Photonics}\ }\textbf {\bibinfo {volume} {11}},\ \bibinfo {pages}
		{149} (\bibinfo {year} {2017})}\BibitemShut {NoStop}%
	\bibitem [{\citenamefont {Capretti}\ \emph {et~al.}(2015)\citenamefont
		{Capretti}, \citenamefont {Wang}, \citenamefont {Engheta},\ and\
		\citenamefont {Dal~Negro}}]{capretti2015TGHITOENZ}%
	\BibitemOpen
	\bibfield  {author} {\bibinfo {author} {\bibfnamefont {A.}~\bibnamefont
			{Capretti}}, \bibinfo {author} {\bibfnamefont {Y.}~\bibnamefont {Wang}},
		\bibinfo {author} {\bibfnamefont {N.}~\bibnamefont {Engheta}}, \ and\
		\bibinfo {author} {\bibfnamefont {L.}~\bibnamefont {Dal~Negro}},\ }\href@noop
	{} {\bibfield  {journal} {\bibinfo  {journal} {Acs Photonics}\ }\textbf
		{\bibinfo {volume} {2}},\ \bibinfo {pages} {1584} (\bibinfo {year}
		{2015})}\BibitemShut {NoStop}%
	\bibitem [{\citenamefont {Kinsey}\ \emph {et~al.}(2015)\citenamefont {Kinsey},
		\citenamefont {DeVault}, \citenamefont {Kim}, \citenamefont {Ferrera},
		\citenamefont {Shalaev},\ and\ \citenamefont
		{Boltasseva}}]{kinsey2015ENZlinear}%
	\BibitemOpen
	\bibfield  {author} {\bibinfo {author} {\bibfnamefont {N.}~\bibnamefont
			{Kinsey}}, \bibinfo {author} {\bibfnamefont {C.}~\bibnamefont {DeVault}},
		\bibinfo {author} {\bibfnamefont {J.}~\bibnamefont {Kim}}, \bibinfo {author}
		{\bibfnamefont {M.}~\bibnamefont {Ferrera}}, \bibinfo {author} {\bibfnamefont
			{V.}~\bibnamefont {Shalaev}}, \ and\ \bibinfo {author} {\bibfnamefont
			{A.}~\bibnamefont {Boltasseva}},\ }\href@noop {} {\bibfield  {journal}
		{\bibinfo  {journal} {Optica}\ }\textbf {\bibinfo {volume} {2}},\ \bibinfo
		{pages} {616} (\bibinfo {year} {2015})}\BibitemShut {NoStop}%
	\bibitem [{\citenamefont {Alam}\ \emph {et~al.}(2016)\citenamefont {Alam},
		\citenamefont {De~Leon},\ and\ \citenamefont {Boyd}}]{alam2016ENZnonlinear}%
	\BibitemOpen
	\bibfield  {author} {\bibinfo {author} {\bibfnamefont {M.~Z.}\ \bibnamefont
			{Alam}}, \bibinfo {author} {\bibfnamefont {I.}~\bibnamefont {De~Leon}}, \
		and\ \bibinfo {author} {\bibfnamefont {R.~W.}\ \bibnamefont {Boyd}},\
	}\href@noop {} {\bibfield  {journal} {\bibinfo  {journal} {Science}\ }\textbf
		{\bibinfo {volume} {352}},\ \bibinfo {pages} {795} (\bibinfo {year}
		{2016})}\BibitemShut {NoStop}%
	\bibitem [{\citenamefont {Caspani}\ \emph {et~al.}(2016)\citenamefont
		{Caspani}, \citenamefont {Kaipurath}, \citenamefont {Clerici}, \citenamefont
		{Ferrera}, \citenamefont {Roger}, \citenamefont {Kim}, \citenamefont
		{Kinsey}, \citenamefont {Pietrzyk}, \citenamefont {Di~Falco}, \citenamefont
		{Shalaev}, \citenamefont {Boltasseva},\ and\ \citenamefont
		{Faccio}}]{PRLCaspani}%
	\BibitemOpen
	\bibfield  {author} {\bibinfo {author} {\bibfnamefont {L.}~\bibnamefont
			{Caspani}}, \bibinfo {author} {\bibfnamefont {R.~P.~M.}\ \bibnamefont
			{Kaipurath}}, \bibinfo {author} {\bibfnamefont {M.}~\bibnamefont {Clerici}},
		\bibinfo {author} {\bibfnamefont {M.}~\bibnamefont {Ferrera}}, \bibinfo
		{author} {\bibfnamefont {T.}~\bibnamefont {Roger}}, \bibinfo {author}
		{\bibfnamefont {J.}~\bibnamefont {Kim}}, \bibinfo {author} {\bibfnamefont
			{N.}~\bibnamefont {Kinsey}}, \bibinfo {author} {\bibfnamefont
			{M.}~\bibnamefont {Pietrzyk}}, \bibinfo {author} {\bibfnamefont
			{A.}~\bibnamefont {Di~Falco}}, \bibinfo {author} {\bibfnamefont {V.~M.}\
			\bibnamefont {Shalaev}}, \bibinfo {author} {\bibfnamefont {A.}~\bibnamefont
			{Boltasseva}}, \ and\ \bibinfo {author} {\bibfnamefont {D.}~\bibnamefont
			{Faccio}},\ }\href {\doibase 10.1103/PhysRevLett.116.233901} {\bibfield
		{journal} {\bibinfo  {journal} {Phys. Rev. Lett.}\ ,\ \bibinfo {pages}
			{233901}} (\bibinfo {year} {2016})}\BibitemShut {NoStop}%
	\bibitem [{\citenamefont {Clerici}\ \emph {et~al.}(2017)\citenamefont
		{Clerici}, \citenamefont {Kinsey}, \citenamefont {DeVault}, \citenamefont
		{Kim}, \citenamefont {Carnemolla}, \citenamefont {Caspani}, \citenamefont
		{Shaltout}, \citenamefont {Faccio}, \citenamefont {Shalaev}, \citenamefont
		{Boltasseva} \emph {et~al.}}]{clericiNC}%
	\BibitemOpen
	\bibfield  {author} {\bibinfo {author} {\bibfnamefont {M.}~\bibnamefont
			{Clerici}}, \bibinfo {author} {\bibfnamefont {N.}~\bibnamefont {Kinsey}},
		\bibinfo {author} {\bibfnamefont {C.}~\bibnamefont {DeVault}}, \bibinfo
		{author} {\bibfnamefont {J.}~\bibnamefont {Kim}}, \bibinfo {author}
		{\bibfnamefont {E.~G.}\ \bibnamefont {Carnemolla}}, \bibinfo {author}
		{\bibfnamefont {L.}~\bibnamefont {Caspani}}, \bibinfo {author} {\bibfnamefont
			{A.}~\bibnamefont {Shaltout}}, \bibinfo {author} {\bibfnamefont
			{D.}~\bibnamefont {Faccio}}, \bibinfo {author} {\bibfnamefont
			{V.}~\bibnamefont {Shalaev}}, \bibinfo {author} {\bibfnamefont
			{A.}~\bibnamefont {Boltasseva}},  \emph {et~al.},\ }\href@noop {} {\bibfield
		{journal} {\bibinfo  {journal} {Nature communications}\ }\textbf {\bibinfo
			{volume} {8}},\ \bibinfo {pages} {15829} (\bibinfo {year}
		{2017})}\BibitemShut {NoStop}%
	\bibitem [{\citenamefont {Yang}\ \emph {et~al.}(2017)\citenamefont {Yang},
		\citenamefont {Kelley}, \citenamefont {Sachet}, \citenamefont {Campione},
		\citenamefont {Luk}, \citenamefont {Maria}, \citenamefont {Sinclair},\ and\
		\citenamefont {Brener}}]{yang2017femtosecond}%
	\BibitemOpen
	\bibfield  {author} {\bibinfo {author} {\bibfnamefont {Y.}~\bibnamefont
			{Yang}}, \bibinfo {author} {\bibfnamefont {K.}~\bibnamefont {Kelley}},
		\bibinfo {author} {\bibfnamefont {E.}~\bibnamefont {Sachet}}, \bibinfo
		{author} {\bibfnamefont {S.}~\bibnamefont {Campione}}, \bibinfo {author}
		{\bibfnamefont {T.~S.}\ \bibnamefont {Luk}}, \bibinfo {author} {\bibfnamefont
			{J.-P.}\ \bibnamefont {Maria}}, \bibinfo {author} {\bibfnamefont {M.~B.}\
			\bibnamefont {Sinclair}}, \ and\ \bibinfo {author} {\bibfnamefont
			{I.}~\bibnamefont {Brener}},\ }\href@noop {} {\bibfield  {journal} {\bibinfo
			{journal} {Nature Photonics}\ }\textbf {\bibinfo {volume} {11}},\ \bibinfo
		{pages} {390} (\bibinfo {year} {2017})}\BibitemShut {NoStop}%
	\bibitem [{\citenamefont {Carnemolla}\ \emph {et~al.}(2018)\citenamefont
		{Carnemolla}, \citenamefont {Caspani}, \citenamefont {DeVault}, \citenamefont
		{Clerici}, \citenamefont {Vezzoli}, \citenamefont {Bruno}, \citenamefont
		{Shalaev}, \citenamefont {Faccio}, \citenamefont {Boltasseva},\ and\
		\citenamefont {Ferrera}}]{carnemolla2018degenerate}%
	\BibitemOpen
	\bibfield  {author} {\bibinfo {author} {\bibfnamefont {E.~G.}\ \bibnamefont
			{Carnemolla}}, \bibinfo {author} {\bibfnamefont {L.}~\bibnamefont {Caspani}},
		\bibinfo {author} {\bibfnamefont {C.}~\bibnamefont {DeVault}}, \bibinfo
		{author} {\bibfnamefont {M.}~\bibnamefont {Clerici}}, \bibinfo {author}
		{\bibfnamefont {S.}~\bibnamefont {Vezzoli}}, \bibinfo {author} {\bibfnamefont
			{V.}~\bibnamefont {Bruno}}, \bibinfo {author} {\bibfnamefont {V.~M.}\
			\bibnamefont {Shalaev}}, \bibinfo {author} {\bibfnamefont {D.}~\bibnamefont
			{Faccio}}, \bibinfo {author} {\bibfnamefont {A.}~\bibnamefont {Boltasseva}},
		\ and\ \bibinfo {author} {\bibfnamefont {M.}~\bibnamefont {Ferrera}},\
	}\href@noop {} {\bibfield  {journal} {\bibinfo  {journal} {Optical Materials
				Express}\ }\textbf {\bibinfo {volume} {8}},\ \bibinfo {pages} {3392}
		(\bibinfo {year} {2018})}\BibitemShut {NoStop}%
	\bibitem [{\citenamefont {Niu}\ \emph {et~al.}(2018)\citenamefont {Niu},
		\citenamefont {Hu}, \citenamefont {Chu},\ and\ \citenamefont
		{Gong}}]{niu2018}%
	\BibitemOpen
	\bibfield  {author} {\bibinfo {author} {\bibfnamefont {X.}~\bibnamefont
			{Niu}}, \bibinfo {author} {\bibfnamefont {X.}~\bibnamefont {Hu}}, \bibinfo
		{author} {\bibfnamefont {S.}~\bibnamefont {Chu}}, \ and\ \bibinfo {author}
		{\bibfnamefont {Q.}~\bibnamefont {Gong}},\ }\href@noop {} {\bibfield
		{journal} {\bibinfo  {journal} {Advanced Optical Materials}\ }\textbf
		{\bibinfo {volume} {6}},\ \bibinfo {pages} {1701292} (\bibinfo {year}
		{2018})}\BibitemShut {NoStop}%
	\bibitem [{\citenamefont {Wood}\ \emph {et~al.}(2018)\citenamefont {Wood},
		\citenamefont {Campione}, \citenamefont {Parameswaran}, \citenamefont {Luk},
		\citenamefont {Wendt}, \citenamefont {Serkland},\ and\ \citenamefont
		{Keeler}}]{wood2018}%
	\BibitemOpen
	\bibfield  {author} {\bibinfo {author} {\bibfnamefont {M.~G.}\ \bibnamefont
			{Wood}}, \bibinfo {author} {\bibfnamefont {S.}~\bibnamefont {Campione}},
		\bibinfo {author} {\bibfnamefont {S.}~\bibnamefont {Parameswaran}}, \bibinfo
		{author} {\bibfnamefont {T.~S.}\ \bibnamefont {Luk}}, \bibinfo {author}
		{\bibfnamefont {J.~R.}\ \bibnamefont {Wendt}}, \bibinfo {author}
		{\bibfnamefont {D.~K.}\ \bibnamefont {Serkland}}, \ and\ \bibinfo {author}
		{\bibfnamefont {G.~A.}\ \bibnamefont {Keeler}},\ }\href@noop {} {\bibfield
		{journal} {\bibinfo  {journal} {Optica}\ }\textbf {\bibinfo {volume} {5}},\
		\bibinfo {pages} {233} (\bibinfo {year} {2018})}\BibitemShut {NoStop}%
	\bibitem [{\citenamefont {Reshef}\ \emph {et~al.}(2019)\citenamefont {Reshef},
		\citenamefont {De~Leon}, \citenamefont {Alam},\ and\ \citenamefont
		{Boyd}}]{boyd2019NatRev}%
	\BibitemOpen
	\bibfield  {author} {\bibinfo {author} {\bibfnamefont {O.}~\bibnamefont
			{Reshef}}, \bibinfo {author} {\bibfnamefont {I.}~\bibnamefont {De~Leon}},
		\bibinfo {author} {\bibfnamefont {M.~Z.}\ \bibnamefont {Alam}}, \ and\
		\bibinfo {author} {\bibfnamefont {R.~W.}\ \bibnamefont {Boyd}},\ }\href@noop
	{} {\bibfield  {journal} {\bibinfo  {journal} {Nature Reviews Materials}\ ,\
			\bibinfo {pages} {1}} (\bibinfo {year} {2019})}\BibitemShut {NoStop}%
	\bibitem [{\citenamefont {Shaltout}\ \emph {et~al.}(2019)\citenamefont
		{Shaltout}, \citenamefont {Shalaev},\ and\ \citenamefont
		{Brongersma}}]{amr2019}%
	\BibitemOpen
	\bibfield  {author} {\bibinfo {author} {\bibfnamefont {A.~M.}\ \bibnamefont
			{Shaltout}}, \bibinfo {author} {\bibfnamefont {V.~M.}\ \bibnamefont
			{Shalaev}}, \ and\ \bibinfo {author} {\bibfnamefont {M.~L.}\ \bibnamefont
			{Brongersma}},\ }\href {\doibase 10.1126/science.aat3100} {\bibfield
		{journal} {\bibinfo  {journal} {Science}\ }\textbf {\bibinfo {volume} {364}}
		(\bibinfo {year} {2019}),\ 10.1126/science.aat3100},\ \Eprint
	{http://arxiv.org/abs/https://science.sciencemag.org/content/364/6441/eaat3100.full.pdf}
	{https://science.sciencemag.org/content/364/6441/eaat3100.full.pdf}
	\BibitemShut {NoStop}%
	\bibitem [{\citenamefont {Vezzoli}\ \emph {et~al.}(2018)\citenamefont
		{Vezzoli}, \citenamefont {Bruno}, \citenamefont {DeVault}, \citenamefont
		{Roger}, \citenamefont {Shalaev}, \citenamefont {Boltasseva}, \citenamefont
		{Ferrera}, \citenamefont {Clerici}, \citenamefont {Dubietis},\ and\
		\citenamefont {Faccio}}]{VezzoliTime}%
	\BibitemOpen
	\bibfield  {author} {\bibinfo {author} {\bibfnamefont {S.}~\bibnamefont
			{Vezzoli}}, \bibinfo {author} {\bibfnamefont {V.}~\bibnamefont {Bruno}},
		\bibinfo {author} {\bibfnamefont {C.}~\bibnamefont {DeVault}}, \bibinfo
		{author} {\bibfnamefont {T.}~\bibnamefont {Roger}}, \bibinfo {author}
		{\bibfnamefont {V.~M.}\ \bibnamefont {Shalaev}}, \bibinfo {author}
		{\bibfnamefont {A.}~\bibnamefont {Boltasseva}}, \bibinfo {author}
		{\bibfnamefont {M.}~\bibnamefont {Ferrera}}, \bibinfo {author} {\bibfnamefont
			{M.}~\bibnamefont {Clerici}}, \bibinfo {author} {\bibfnamefont
			{A.}~\bibnamefont {Dubietis}}, \ and\ \bibinfo {author} {\bibfnamefont
			{D.}~\bibnamefont {Faccio}},\ }\href {\doibase
		10.1103/PhysRevLett.120.043902} {\bibfield  {journal} {\bibinfo  {journal}
			{Phys. Rev. Lett.}\ }\textbf {\bibinfo {volume} {120}},\ \bibinfo {pages}
		{043902} (\bibinfo {year} {2018})}\BibitemShut {NoStop}%
	\bibitem [{\citenamefont {Khurgin}\ and\ \citenamefont
		{Kinsey}(2019)}]{khurgin2019}%
	\BibitemOpen
	\bibfield  {author} {\bibinfo {author} {\bibfnamefont {J.~B.}\ \bibnamefont
			{Khurgin}}\ and\ \bibinfo {author} {\bibfnamefont {N.}~\bibnamefont
			{Kinsey}},\ }\href@noop {} {\bibfield  {journal} {\bibinfo  {journal} {arXiv
				preprint arXiv:1906.04849}\ } (\bibinfo {year} {2019})}\BibitemShut {NoStop}%
	\bibitem [{\citenamefont {Pendry}(2008)}]{pendry2008time}%
	\BibitemOpen
	\bibfield  {author} {\bibinfo {author} {\bibfnamefont {J.}~\bibnamefont
			{Pendry}},\ }\href@noop {} {\bibfield  {journal} {\bibinfo  {journal}
			{Science}\ }\textbf {\bibinfo {volume} {322}},\ \bibinfo {pages} {71}
		(\bibinfo {year} {2008})}\BibitemShut {NoStop}%
	\bibitem [{\citenamefont {Sivan}\ and\ \citenamefont
		{Pendry}(2011)}]{PhysRevAPendry}%
	\BibitemOpen
	\bibfield  {author} {\bibinfo {author} {\bibfnamefont {Y.}~\bibnamefont
			{Sivan}}\ and\ \bibinfo {author} {\bibfnamefont {J.~B.}\ \bibnamefont
			{Pendry}},\ }\href {\doibase 10.1103/PhysRevA.84.033822} {\bibfield
		{journal} {\bibinfo  {journal} {Phys. Rev. A}\ }\textbf {\bibinfo {volume}
			{84}},\ \bibinfo {pages} {033822} (\bibinfo {year} {2011})}\BibitemShut
	{NoStop}%
	\bibitem [{\citenamefont {Palomba}\ \emph {et~al.}(2012)\citenamefont
		{Palomba}, \citenamefont {Zhang}, \citenamefont {Park}, \citenamefont
		{Bartal}, \citenamefont {Yin},\ and\ \citenamefont
		{Zhang}}]{palomba2012optical}%
	\BibitemOpen
	\bibfield  {author} {\bibinfo {author} {\bibfnamefont {S.}~\bibnamefont
			{Palomba}}, \bibinfo {author} {\bibfnamefont {S.}~\bibnamefont {Zhang}},
		\bibinfo {author} {\bibfnamefont {Y.}~\bibnamefont {Park}}, \bibinfo {author}
		{\bibfnamefont {G.}~\bibnamefont {Bartal}}, \bibinfo {author} {\bibfnamefont
			{X.}~\bibnamefont {Yin}}, \ and\ \bibinfo {author} {\bibfnamefont
			{X.}~\bibnamefont {Zhang}},\ }\href@noop {} {\bibfield  {journal} {\bibinfo
			{journal} {Nature materials}\ }\textbf {\bibinfo {volume} {11}},\ \bibinfo
		{pages} {34} (\bibinfo {year} {2012})}\BibitemShut {NoStop}%
	\bibitem [{\citenamefont {Rao}\ \emph {et~al.}(2015)\citenamefont {Rao},
		\citenamefont {Lyons}, \citenamefont {Roger}, \citenamefont {Clerici},
		\citenamefont {Zheludev},\ and\ \citenamefont {Faccio}}]{rao2015geometries}%
	\BibitemOpen
	\bibfield  {author} {\bibinfo {author} {\bibfnamefont {S.~M.}\ \bibnamefont
			{Rao}}, \bibinfo {author} {\bibfnamefont {A.}~\bibnamefont {Lyons}}, \bibinfo
		{author} {\bibfnamefont {T.}~\bibnamefont {Roger}}, \bibinfo {author}
		{\bibfnamefont {M.}~\bibnamefont {Clerici}}, \bibinfo {author} {\bibfnamefont
			{N.~I.}\ \bibnamefont {Zheludev}}, \ and\ \bibinfo {author} {\bibfnamefont
			{D.}~\bibnamefont {Faccio}},\ }\href@noop {} {\bibfield  {journal} {\bibinfo
			{journal} {Scientific reports}\ }\textbf {\bibinfo {volume} {5}},\ \bibinfo
		{pages} {15399} (\bibinfo {year} {2015})}\BibitemShut {NoStop}%
\end{thebibliography}
\end{document}